\documentstyle[12pt,aasms4,psfig]{article}

\lefthead{E.~Kuulkers et al.}
\righthead{Dips in GRO\,J1655$-$40 and 4U\,1630$-$47}

\begin{document}

\title{Absorption dips in the light curves of GRO\,J1655$-$40 and 
4U\,1630$-$47 during outburst}

\author{Erik Kuulkers\altaffilmark{1},
	Rudy Wijnands\altaffilmark{2},
        Tomaso Belloni\altaffilmark{2},
	Mariano M\'endez\altaffilmark{2,3},
        Michiel van der Klis\altaffilmark{2},
        Jan van Paradijs\altaffilmark{2,4}}

\altaffiltext{1}{Astrophysics, University of Oxford, Nuclear and
       Astrophysics Laboratory, Keble Road, Oxford OX1 3RH, United
       Kingdom, erik@astro.ox.ac.uk}

\altaffiltext{2}{Astronomical Institute ``Anton Pannekoek'',
       University of Amsterdam and Center for High-Energy Astrophysics,
       Kruislaan 403, NL-1098 SJ Amsterdam, the Netherlands, rudy@astro.uva.nl,
       tmb@astro.uva.nl, mariano@astro.uva.nl, michiel@astro.uva.nl, 
       jvp@astro.uva.nl}

\altaffiltext{3}{Facultad de Ciencias Astron\'omicas y Geof\'{\i}sicas, 
       Universidad Nacional de La Plata, Paseo del Bosque S/N, 
       1900 La Plata, Argentina}

\altaffiltext{4}{Physics Department, University of Alabama in Huntsville,
       Huntsville, AL 35899, USA}

\begin{abstract}

Using the RXTE PCA we discovered deep dips in the X-ray light curves 
of the black-hole candidates GRO\,J1655$-$40 and 4U\,1630$-$47 during 
outburst. Similar kind of dips for GRO\,J1655$-$40 were found in 90\,s 
measurements of the RXTE ASM during the same outburst. The duration of the 
dips in both sources is in the order of minutes. The 
occurrences of the dips observed with the RXTE PCA 
and ASM in GRO\,J1655$-$40 are consistent with the optically determined 
orbital period, and were found between photometric orbital phases 0.72 and 
0.86. This constitutes the first evidence for orbital variations in X-rays for 
GRO\,J1655$-$40. The PCA data indicate that an absorbing medium is 
responsible for these dips. The X-ray spectra during the 
dips can be best described by a heavily absorbed component and an unabsorbed
component. In the case of GRO\,J1655$-$40 we are able constrain
the extent of the absorbing medium and the central X-ray source.

\end{abstract}


\section{Introduction}

X-ray intensity dips caused by an intervening medium have now been 
found in the light curves of various low-mass and high-mass X-ray binaries
(e.g., Parmar \&\ White \cite{pw88}, Marshall et al.\ \cite{mmp93}, 
Saraswat et al.\ \cite{sym96}, and references therein).
During the majority of these dips the X-ray spectra harden, which is 
indicative of photoelectric absorption of radiation from the central
X-ray source. However, a simple neutral and uniform medium which absorbs the 
emission does not fit the X-ray spectra. Instead, the spectra reveal
an excess flux at low energies (typically $\lesssim$4\,keV) compared to that 
expected from the amount of absorption estimated from data above $\sim$4\,keV.

In this {\it Letter} we report on such dips seen in the 
X-ray light curves of the black-hole (candidate) soft X-ray transients 
GRO\,J1655$-$40 and 4U\,1630$-$47 during their 1996/1997 and 1996 outbursts, 
respectively, as obtained with the {\it Proportional Counter Array} (PCA) and
the {\it All Sky Monitor} (ASM) on board the 
{\it Rossi X-ray Timing Explorer} (RXTE). 

GRO\,J1655$-$40 was discovered during an outburst in 1994 
and has since shown irregular outburst activity
(e.g., Zhang et al.\ \cite{zes97}). 
Dynamical measurements suggest the compact star in GRO\,J1655$-$40 
is a black hole, with a mass of $\sim$7\,M$_{\odot}$ 
(Orosz \&\ Bailyn \cite{ob97}, van der Hooft et al.\ \cite{hha97}). 
Recently, Ueda et al.\ (\cite{uit97}) found evidence for a $\sim$5.7\,hr dip 
down to $\sim$25\%\ of the out-of-dip intensity in the ASCA light curve of 
GRO\,J1655$-$40, with no clear evidence of spectral hardening. 
The X-ray dips reported here are considerably deeper (down to $\sim$8\%\ of the 
out-of-dip intensity) and have much shorter duration ($\sim$minutes).
A preliminary announcement of the dips in GRO\,J1655$-$40 observed by RXTE has 
already been given by Kuulkers et al.\ (\cite{kbm97}).

4U\,1630$-$47 has shown outbursts every $\sim$600 days since at least 1969
(e.g.\ Kuulkers et al.\ \cite{kpk97}).
The nature of the compact star in 4U\,1630$-$47 is unknown.
Its X-ray spectral (e.g.\ Barret, McClintock, \&\ Grindlay \cite{bmg96})
and X-ray timing (Kuulkers, van der Klis \&\ Parmar \cite{kkp97}) properties 
during outburst suggest it is a black-hole.
Recently, Kuulkers et al.\ (\cite{kkp97}; \cite{kpk97}) pointed out 
similarities in the X-ray behavior between 4U\,1630$-$47 and GRO\,J1655$-$40, 
and postulated that they are similar kind of systems.

\section{Observations and Analysis}

The RXTE PCA (Bradt, Rothschild, \&\ Swank \cite{brs93})
performed a public Target of Opportunity Observations of 4U\,1630$-$47 on
1996 May 3 20:49--22:41~UTC, and of GRO\,J1655$-$40 
on 1997 February 26 19:34--23:30~UTC. 
The data were collected with a time resolution of 16\,s (129 photon energy
channels, covering 2.0--60\,keV) and 125\,$\mu$s 
(3 energy channels, covering 2.0--5.0--8.7--13.0\,keV). The 16\,s data
were used in the X-ray spectral analysis. From the high time resolution
data we constructed light curves by rebinning the ``raw'' 
(i.e., not corrected for dead time) 
background subtracted count rates to a time resolution of 
0.25\,s, and hardness values using count rate ratios binned 
to a time resolution of 0.25\,s (GRO\,J1655$-$40) or 0.5\,s (4U\,1630$-$47).
All X-ray spectral fits were performed in the range 2--30\,keV (out of dip)
or 2--20\,keV (dip), and a 2\%\ uncertainty was included in the data to 
account for uncertainties in the PCA response matrix 
(see e.g., Belloni et al.\ \cite{bmk97}).
The X-ray spectra were corrected for background and dead time.
Errors quoted for the spectral parameters were determined 
using $\Delta\chi^2$=2.706 (90\%\ confidence).

The ASM (Levine et al.\ \cite{lbcj96}) on board RXTE 
scans the sky in series of 90\,s dwells in three energy bands, 1.5--3, 3--5, 
and 5--12\,keV. Due to satellite motion and a $\sim$40\%\ duty cycle, any given 
source is scanned 5--10 times per day. For our analysis we used
the results provided by the RXTE ASM team, covering the period from 1996 
February 21 to 1997 Jun 19.

\section{Results}

\subsection{Intensity and hardness variations}

The 2.0--13.0\,keV count rates of GRO\,J1655$-$40 and 4U\,1630$-$47 were
generally $\sim$11\,500\,cts\,s$^{-1}$ and $\sim$3000\,cts\,s$^{-1}$, 
respectively. However, both sources showed sharp deep drops
in their count rates down to $\sim$1000\,cts\,s$^{-1}$,
lasting $\sim$50--60\,s for GRO\,J1655$-$40 (two events), and 
$\sim$140\,s for 4U\,1630$-$47 (one event), preceded and followed by short 
dips (Fig.~1, upper panel). We will refer to the out-of-dip count rate as 
the ``persistent emission'' (PE).

The fall time scales of the main dips of GRO\,J1655$-$40 are 
2--4\,s, while the rise time scales are 3--5\,s. 
The hardness curve (Fig.~1, lower left panel) and 
corresponding hardness vs.\ intensity diagram 
(HID, Fig.~2, upper panel) of GRO\,J1655$-$40 
show that as the intensity drops the X-ray 
spectrum becomes much harder, until a certain threshold count rate of 
$\sim$2000\,cts\,s$^{-1}$ is reached.
During the last part of the fall the spectrum softens again 
to almost the same value as the PE level. The reverse 
behaviour is seen during the rise of the dips.
In the HID the source always follows the same loop, also 
during the small pre- and after-dips and during the short spike 
in the second dip.

The fall and rise time scale of the main dip of 4U\,1630$-$47 are $\sim$3\,s.
During its main dip several small flares occurred.
The light curve and hardness behavior (Fig.~1, lower right panel; 
Fig.~2, lower panel) of 4U\,1630$-$47 resembles that of GRO\,J1655$-$40. 
Below a certain threshold count rate the hardness is positively 
correlated with intensity.
However, the change in hardness between the persistent and threshold count rate 
is much smaller than that observed in GRO\,J1655$-$40. 
Below the threshold count rate the hardness drops below the 
persistent value.

In Fig.~3 we plot the RXTE ASM light curve of the individual 90\,s dwells
for GRO\,J1655$-$40.
12 clear drops in the 2-12\,keV intensity can be seen down by $\sim$25--95\%. 
During these dips the spectral hardness increases. 
The occurrences of the ASM and PCA dips (Table~1) are best fit with a period of 
2.6213$\pm$0.0005 days (1$\sigma$).
All these dips occurred between photometric orbital phases 0.72 and 0.86 
(Table~1). 

\subsection{Energy spectra}

We constructed a series of $\sim$200\,sec X-ray spectra just outside the 
dips, and individual 16\,sec spectra at the bottom of the dips.
Because of the 16\,s time resolution we could not 
accumulate spectra during the rise and fall of the dips.
The average count rates during the dip spectra are all below the 
threshold count rate.

Fits to the PE with simple single-component models did not 
give acceptable results. We therefore used the model usually employed for 
black-hole candidates, i.e., a disk-black body (DBB) plus a power law
(Mitsuda et al.\ \cite{mxx84}), which gave values of $\chi^2_{\nu}$$\lesssim$2.
Using the mean spectral fits (Table~2) we find that the persistent 2--30\,keV 
X-ray flux was $\sim$2.5$\times$10$^{-8}$ and 
$\sim$5.8$\times$10$^{-9}$\,erg\,cm$^{-2}$\,s$^{-1}$, for GRO\,J1655$-$40 
and 4U\,1630$-$47, respectively.

The spectral hardening between the PE count rate and the 
threshold value suggests that absorption is involved in the process giving 
rise to the dips. To verify this, we also made color-color 
diagrams (CDs, see Fig.~3). The data 
points for the two sources during the dips move from the crowded regions 
(PE) to the upper right part and then bend back 
to the lower left part of the CD (in the case of 4U\,1630$-$47 the data 
points bend back almost immediately). The bend-over point corresponds to the 
threshold count rate. We also plotted the calculated hardness values for 
the DBB and power-law component separately for a range of values of the 
inner disk temperature, $T_{\rm in}$, and of the power-law index, 
$\Gamma$, with interstellar absorption column density, $N_{\rm H_{\rm int}}$.
fixed to the persistent value (dashed lines in Fig.~2). 
With our choice of hardness ratios any linear 
combination of the two components has to lie between these two lines. 
Only in the CD of GRO\,J1655$-$40 part of the loop lies 
below the DBB line; this requires an increase in the absorption column.

Homogeneous absorption of the persistent DBB and power-law components by cold 
material did not fit the dip spectra well, especially at low 
energies. The observed flux below 
$\sim$6\,keV is much in excess from that expected in this model.
We performed simple fits to the spectra by modelling 
this low energy excess either as a separate component 
(power law or black body), 
or by partial absorption of the persistent components 
(e.g.\ Marshall et al.\ \cite{mmp93}).
We find that the persistent power-law component is absent during 
the dips. We focus on the fit results 
to the spectra at the bottom of the dips (Table~3) 
to investigate the low-energy 
excess, since there the absorption is expected to be largest. 
We fixed $N_{\rm H_{\rm int}}$ to the values derived from the
PE fits, except when using the partial covering absorption 
model for 4U\,1630$-$47 because this lead to an unstable fit.
When using a power law to model the low-energy excess plus an absorbed DBB for 
the dip spectrum of 4U\,1630$-$47, the inner disk radius, $R_{\rm in}$, 
could not be constrained and was, therefore, fixed to its PE
value. 

The fits show that the shape and strength of the low-energy excess 
is very similar for both sources, and its contribution is only 6--7\%\
of the PE flux. When the low-energy excess is modeled as 
a separate component, the DBB parameters are consistent with the 
PE values. When a partial covering model is used
to describe the spectra, we get somewhat lower values for 
$T_{\rm in}$ with respect to their PE values.
Depending on the model we find that the absorption of the 
DBB component increased up to $\sim$25--200$\times$10$^{22}$ and
$\sim$35--300$\times$10$^{22}$\,cm$^{-2}$
at the lowest mean dip intensities in GRO\,J1655$-$40 and 4U\,1630$-$47, 
respectively.

To see if we can qualitatively reproduce the observed shapes of the 
HID and CD, we calculated several sequences of X-ray spectra and determined
intensity and hardness values. 
The out-of-dip spectrum was modeled by the persistent DBB component 
(only subject to interstellar absorption) as given in Table~2, plus the 
low-energy excess contribution modeled by a black body, as given in Table~3.
In the dip we linearly increased the absorption of the persistent 
DBB component from zero up to 150 and 
300$\times$10$^{22}$\,cm$^{-2}$, for 
GRO\,J1655$-$40 and 4U\,1630$-$47 respectively, fixing the rest of the 
parameters to those given by the out-of-dip spectrum.
The results are plotted in Fig.~2, and show that a gradual increase in 
absorption of the DBB component can reproduce the observed dip behavior.

\section{Discussion}

\subsection{RXTE PCA and ASM dips}

We have discovered short-term ($\sim$minutes) 
X-ray dipping behaviour of GRO\,J1655$-$40 and 4U\,1630$-$47, down to
$\sim$8\%\ and $\sim$30\%\ of the out-of-dip intensity, respectively.
We found similar intensity drops in GRO\,J1655$-$40 
during 90\,s measurements with the RXTE ASM. 
The duration of the dips is of the same order of those found in Cyg\,X-1 
and Her\,X-1 (e.g., Kitamoto et al.\ \cite{kmt84},
Leahy \cite{l97}), but is shorter than
typically seen in LMXB dip sources (e.g., Parmar \&\ White 1988).

The best fit period of the
occurrence of the dips observed in GRO\,J1655$-$40 
is consistent with the optical
period of the system (Orosz \&\ Bailyn 1997, van der Hooft et al.\ 
\cite{hha97}). This therefore constitutes the first evidence of 
the orbital period in GRO\,J1655$-$40 in X-rays. All these dips occurred 
between photometric orbital phases 0.72 and 0.86.
We note that the X-ray dip in GRO\,J1655$-$40 seen with ASCA 
(Ueda et al.\ 1997) occurred near photometric orbital phase 0.6.
The phasing of the occurrence of the X-ray dips is very similar to that 
observed in the low-mass X-ray binary dip sources 
(e.g., Parmar \&\ White \cite{pw88}),
which suggests a similar origin for the cause of the dips.
The inclination of such sources which show only dips and no eclipses
are in the range 60--75$^{\circ}$ (e.g., Frank , King \&\ Lasota \cite{fkl87}). 
The inclination inferred for GRO\,J1655$-$40
(Orosz \&\ Bailyn \cite{ob97}, van der Hooft et al.\ \cite{hha97}) is 
in agreement with this. Since the dip behavior of 4U\,1630$-$47 
is so similar to that seen in GRO\,J1655$-$40, we propose that 
4U\,1630$-$47 is also seen at a relatively high inclination, i.e.\ 
60--75$^{\circ}$. 

\subsection{An absorbing medium}

The observed dips are caused by an intervening 
medium absorbing the out-of-dip (persistent) emission (PE). 
A simple neutral and uniform medium which heavily absorbs the PE,
however, does not describe the spectra during the dips very well. 
The dip spectra can be best described by a 
heavily absorbed persistent component, plus a low-energy component which 
typically appears below $\sim$6\,keV. Such an extra low-energy component has 
also been reported by Ueda et al.\ (1997) in GRO\,J1655$-$40 during their dip 
observed with ASCA, and has been seen during dips and 
eclipses in other low-mass and high-mass X-ray binaries 
(Sect.~1). We modeled this so-called low-energy excess as a power law or a 
black body (subject only to interstellar absorption), or alternatively
by partial covering absorption of the 
persistent components.
We find that the maximum absorption of the persistent components 
at the bottom of the dips ($\sim$25--300$\times$10$^{22}$\,cm$^{-2}$) 
is comparable to that found during the long dip in GRO\,J1655$-$40 observed 
with ASCA (Ueda et al.\ 1997) and the bottom of deep dips and/or 
eclipses of other binaries.

Several models have been proposed to explain 
the low-energy excess during dips and/or eclipses (e.g., Marshall et al.\ 
\cite{mmp93}). These include either a separate component which is not 
affected by the variable heavy absorption, or partial covering by a clumpy 
medium. The former has been proposed to be due to a scattering halo
(e.g., Kitamoto, Miyamoto \&\ Yamamoto \cite{kmy89}), or 
ionized absorber models, in which the absorbing medium is sufficiently ionized
to reduce the soft X-ray absorption (e.g., Parmar \&\ White 1988, 
Marshall et al.\ 1993). We note that
Ueda et al.\ (1997) reported iron absorption lines during ASCA 
out-of-dip spectra of GRO\,J1655$-$40 which suggests the presence of 
highly ionized plasma. Alternatively, the low-energy excess may be the 
result of accumulating spectra over time scales longer than the intrinsic 
time scale for variability during the dips (e.g., Parmar et al.\ \cite{pwg86}).

Our spectral fits indicate that during the dips 
the persistent power-law component is absent. 
This suggests that the region where the bulk of the X-rays originate
is not only subject to absorption, but that also the power-law emitting
region is blocked from our view.
Interaction of the inflowing gas stream from the secondary with the 
outer edge of the disk may cause a thickening of the outer edge and/or
material above the disk in the expected 0.6--0.0 phase range
(e.g., Parmar \&\ White \cite{pw88}) and may well cause this absorption and
shadowing. 

Recently, Greiner et al.\ (\cite{gmr96}) showed that the Galatic superluminal 
source GRS\,1915+105 had ``dipping'' behavior during its outbursts. 
The dipping activity in GRS\,1915+105 is much more complex than
that seen in our observations of GRO\,J1655$-$40 and 4U\,1630$-$47.
This dipping behavior has been proposed to be due to thermal-viscous 
instabilities in the inner disk (Belloni et al.\ \cite{bmk97}), and
is therefore not related to the dipping behavior we see 
in GRO\,J1655$-$40 and 4U\,1630$-$47.

\subsection{GRO\,J1655$-$40}

For GRO\,J1655$-$40 the system parameters have been well determined
(Orosz \&\ Bailyn 1997, Van der Hooft et al.\ 1997). In this 
system the time scales of the dips imposes constraints on the sizes
of the different emitting and absorbing media.
The fall and rise time ($t_{\rm r,f}$$\sim$3.5\,s) 
constrain the size of the region which is "obscured";
in fact, this gives an upper limit on the size, because the region over 
which the column density increases significantly also has a finite width
(Leahy, Yoshida, \&\ Matsuoka \cite{lym94}). 
The duration of the dips ($t_{\rm dip}$$\sim$55\,sec) 
constrains the size of the absorbing medium itself.

Since $t_{\rm r,f}$$\ll$$t_{\rm dip}$, we may assume that 
the absorbed X-ray source is much smaller than the absorbing medium. 
A medium which crosses a 
point-like central X-ray source may produce irregular X-ray dips, whereas 
crossing an extended region such as an accretion disk corona may produce smoother and 
longer energy independent modulations (e.g., Parmar \&\ White \cite{pw88}). 
If the medium corotates in the binary 
frame and is located at a radius which is smaller than the outer disk radius
($r_{\rm d}$$\sim$0.85\,R$_{\rm L}$ [Orosz \&\ Bailyn 1997], where 
R$_{\rm L}$ is the effective Roche lobe radius of the black hole) the upper
limit on the size of the X-ray emitting region is $\sim$460\,km. If the 
medium corotates with matter in the accretion disk (i.e., with a Kepler 
velocity), the upper limit becomes $\sim$1600\,km. Similar reasoning 
gives an approximate upper limit on the size of the absorbing medium of 
$\sim$3800\,km or $\sim$37\,000\,km, in the case of rotation within the binary 
frame or corotation in the accretion disk.

If partial ionization of material in the disk causes the low-energy 
excess during the dips, one can roughly estimate the location of the
absorbing material (e.g., Parmar et al.\ 1986).
For the material causing the intensity dips to be significantly ionized the 
ionization parameter $\xi=L/nR^2$ (where $L$ is the central source luminosity,
$n$ the gas density of the cloud and 
$R$ the distance from the central source to the clouds) must be larger than 
$\sim$100\,ergs\,cm\,s$^{-1}$ (see Hatchett, Buff \&\ McCray \cite{hbm76}).
Using a typical column density of 
$\sim$100$\times$10$^{22}$\,cm$^{-2}$
at the bottom of the dips,
an out-of-dip luminosity of L$\simeq$3$\times$10$^{37}$\,ergs\,s$^{-1}$
(at a distance of 3.2\,kpc, Hjellming \&\ Rupen 1995), and 
following Parmar et al.\ (\cite{pwg86}), we derive that the material closer than 
$\sim$3$\times$10$^{5}$\,km from the X-ray source will be significantly 
ionized. Assuming the absorbing medium is more or less spherical and has a 
fixed position in the binary system, one gets (e.g., Remillard \&\ Canizares 
\cite{rc84}) $\xi=2\pi Lt_{\rm dip}(N_{\rm H}P_{\rm orb}R)^{-1}\gtrsim 100$, 
and therefore $R$$\lesssim$5000\,km.
So, if ionization plays a role, the absorbing medium should be 
located in the inner part of the disk. 
Frank et al.\ (\cite{fkl87}) suggested 
that the low-energy excess is caused by scattering of X-rays in hot clouds 
which are the result of material from the accretion stream which
skimmed over the disk and formed a ring near the central X-ray source.
Ionization may generate a two-phase medium which produce the dips.
The expected circularization radius of the ring of material in this model 
is rather large for GRO\,J1655$-$40 
($\sim$1$\times$10$^{6}$\,km) so that material in this ring can not be 
easily ionized.

\acknowledgements

We thank Phil Charles and various participants of the 
{\em Microquasar Workshop} (1997, May 1--3, GSFC, Greenbelt, Maryland)
for stimulating discussions. Keith Jahoda is acknowledged for providing 
an up-to-date PCA response matrix.
This work was supported in part by the Netherlands Organization for
Scientific Research (NWO) under grant PGS 78-277 and by the
Netherlands Foundation for research in astronomy (ASTRON).  MM is a
fellow of the Consejo Nacional de Investigaciones Cient\'{\i}ficas y
T\'ecnicas de la Rep\'ublica Argentina. JvP acknowledges support from the 
National Aeronautics and Space Administration through contract NAG5-3269.

\newpage

\section*{Figure captions}

{\bf Figure 1}: The light curves (upper panel) and hardness curves
(lower panel) of GRO\,J1655$-$40 (left panel) and 4U\,1630$-$47 (right panel).
Hardness is defined as the ratio of 
the count rates in the 5.0--13.0\,keV and 2.0--5.0\,keV bands.
The time resolution in all 
panels is 0.25\,s, except for the lower right panel where it is 0.5\,s.
T=0\,s corresponds to 1997 May 3, 21:22:21~UTC for GRO\,J1655$-$40 and 
1996 Feb 26, 20:48:46~UTC for 4U\,1630$-$47. \\

{\bf Figure 2}: Upper panel: hardness-intensity diagrams for GRO\,J1655$-$40 
(0.25\,s time resolution; left) and 4U\,1630$-$47 
(0.5\,s time resolution; right), where hardness is defined as in Fig.~1. 
The line through the data points corresponds to a simple model of the X-ray 
spectra as described in the text.
Lower panel: ratio of the count rates in the
8.7--13.0\,keV and 2.0--5.0\,keV bands ("hard hardness") versus the ratio of 
the count rates in the 5.0--8.7\,keV and 2.0--5.0\,keV bands
("soft hardness") for GRO\,J1655$-$40 (left) and 
4U\,1630$-$47 (right). The data points (1\,s averages) are given by open 
circles. The upper left dashed line is for the power-law component, 
whereas the lower right dashed line is for the disk black-body component.
The line through the data points corresponds to a simple model of the X-ray 
spectra as described in the text.\\

{\bf Figure 3}: RXTE ASM lightcurve of GRO\,J1655$-$40 of data 
from individual dwells of $\sim$90\,s from 1996 Feb 21 (JD\,2\,450\,135) to 
1997 Jun 19 (JD\,2\,450\,618).
Datapoints separated by $<$2\,d have been connected to guide the eye. 
Clearly, deep sharp dips can be seen. The arrows at the bottom denote the dips 
used in the text. Indicated is also the time of the dips observed with the PCA. 

\newpage

\tiny
\begin{table}
\caption{GRO\,J1655$-$40 dip occurrence times}
\begin{tabular}{lclclc}
\tableline\tableline
 Time & $\phi_{\rm orb}$\tablenotemark{a} &  Time & 
$\phi_{\rm orb}$\tablenotemark{a} & 
Time & $\phi_{\rm orb}$\tablenotemark{a} \\
\multicolumn{2}{l}{(JD$-$2\,450\,000)} & 
\multicolumn{2}{l}{(JD$-$2\,450\,000)} & 
\multicolumn{2}{l}{(JD$-$2\,450\,000)} \\
\tableline
 205.0796 & 0.86 & 225.2172 & 0.73 & 487.4261 & 0.75 \\     
 220.0827 & 0.78 & 225.2183 & 0.73 & 487.4909 & 0.78 \\     
 220.0850 & 0.78 & 228.1535 & 0.85 & 505.9143\tablenotemark{b} & 0.81 \\
 222.6122 & 0.74 & 228.1546 & 0.85 & 505.9163\tablenotemark{b} & 0.81 \\
 222.8194 & 0.82 & 254.3466 & 0.85 & 555.7916 & 0.83 \\
 222.8205 & 0.82 & 280.2353 & 0.72 & 555.7927 & 0.83 \\
 222.8681 & 0.84 & 296.2989 & 0.85 & 565.9994 & 0.73 \\
 222.8692 & 0.84 & 314.4044 & 0.75 &          &      \\
\tableline
\multicolumn{6}{l}{\tablenotemark{a}\,\,Photometric orbital phase (Orosz \&\ Bailyn 1997).} \\
\multicolumn{6}{l}{\tablenotemark{b}\,\,RXTE PCA, see Sect.~3.1.} \\
\end{tabular}
\end{table}

\tiny
\begin{table}
\caption{Persistent emission: X-ray spectral fit results\tablenotemark{a}}
\begin{tabular}{ccllllcl}
\tableline\tableline
\multicolumn{1}{c}{Source} & \multicolumn{1}{c}{$\overline{I}$} & 
\multicolumn{1}{c}{$N_{\rm H_{\rm int}}$} & 
\multicolumn{1}{c}{$T_{\rm in}$} & \multicolumn{1}{c}{$R_{\rm in}$} & 
\multicolumn{1}{c}{$\Gamma$} & \multicolumn{1}{c}{A$_{\rm pl}$} & 
\multicolumn{1}{c}{$\chi^2_{\nu}$/dof} \\ 
\multicolumn{1}{c}{~} & \multicolumn{1}{c}{(cts\,s$^{-1}$)} & 
\multicolumn{1}{c}{(10$^{22}$\,cm$^{-2}$)} & \multicolumn{1}{c}{(keV)} & 
\multicolumn{1}{c}{(km)} & \multicolumn{1}{c}{~} & 
\multicolumn{1}{c}{cm$^{-2}$\,s$^{-1}$\,keV$^{-1}$)} & \multicolumn{1}{c}{~} \\
\tableline
GRO\,J1655$-$40 & 10807 & 2.17$\pm$0.11 & 1.110$\pm$0.002 &24.7$\pm$1.0\tablenotemark{b} & 2.48$\pm$0.02 &0.29$\pm$0.01 & 1.6/54 \\
4U\,1630$-$47 & 2759 & 12.3$\pm$0.3 & 1.277$\pm$0.005 &17.9$\pm$0.3\tablenotemark{c} & 3.65$\pm$0.12 & 26$\pm$8 & 2.0/54 \\
\tableline
\multicolumn{8}{l}{\tablenotemark{a}\,\,Model: $e^{-N_{\rm H_{\rm int}}\sigma(E)}$[A$_{\rm pl}E^{-\gamma }+$DBB$(T_{\rm in},R_{\rm in})$], where $N_{\rm H_{\rm int}}$ is the interstellar absorption column density, $E$} \\
\multicolumn{8}{l}{$^{~}$ the photon energy, and DBB = Disk Black-Body model.} \\
\multicolumn{8}{l}{\tablenotemark{b}\,\,Assuming a distance of 3.2\,kpc (Hjellming \&\ Rupen 1995) and inclination of 70$^{\circ}$ (Orosz \&\ Bailyn 1997; Van} \\
\multicolumn{8}{l}{$^{~}$ der Hooft et al.\ 1997).} \\
\multicolumn{8}{l}{\tablenotemark{c}\,\,In units of ${1\over{\sqrt{\cos i}}}\cdot{\rm d}_{10}$, where $i$ is the inclination of the accretion disk and d$_{10}$ the distance in units of 10\,kpc.} \\
\end{tabular}
\end{table}

\tiny
\begin{table}
\caption{Dips: X-ray spectral fit results}
\begin{tabular}{cllclllc}
\tableline\tableline
\multicolumn{1}{c}{~} &
\multicolumn{7}{c}{power law + absorbed DBB\tablenotemark{a}} \\
\multicolumn{1}{c}{dip\tablenotemark{b}} & 
\multicolumn{1}{c}{$N_{\rm H_{\rm int}}$} & 
\multicolumn{1}{c}{$\gamma$} & \multicolumn{1}{c}{A$_{\rm pl}$} & 
\multicolumn{1}{c}{$N_{\rm H}$} & 
\multicolumn{1}{c}{$T_{\rm in}$} & \multicolumn{1}{c}{$R_{\rm in}$\tablenotemark{c}} & 
\multicolumn{1}{c}{$\chi^2_{\nu}$/dof} \\
\multicolumn{1}{c}{~} & \multicolumn{1}{c}{(10$^{22}$\,cm$^{-2}$)} &
\multicolumn{1}{c}{~} & \multicolumn{1}{c}{(cm$^{-2}$\,s$^{-1}$\,keV$^{-1}$)} & 
\multicolumn{1}{c}{(10$^{22}$\,cm$^{-2}$)} & \multicolumn{1}{c}{(keV)} &
\multicolumn{1}{c}{(km)} & \multicolumn{1}{c}{~} \\
\tableline
GRO\,J1655$-$40 & 2.17\tablenotemark{d} & 4.6$^{+0.8}_{-0.2}$ & 34$^{+35}_{-6}$ & 27$^{+26}_{-18}$ & 1.1$\pm$0.2 & 7$^{+18}_{-3}$ & 2.0/41 \\
4U\,1630$-$47 & 12.3\tablenotemark{d} & 4.5$\pm$0.1 & 73$^{+10}_{-18}$ & 34$\pm$27 & 0.9$\pm$0.1 & 17.9\tablenotemark{d} & 1.2/42 \\
\tableline
\multicolumn{1}{l}{~} & 
\multicolumn{7}{c}{black body (BB) + absorbed DBB\tablenotemark{e}}\\
\multicolumn{1}{c}{dip} & 
\multicolumn{1}{c}{$N_{\rm H_{\rm int}}$} & 
\multicolumn{1}{c}{$T_{\rm bb}$} & \multicolumn{1}{c}{$R_{\rm bb}$\tablenotemark{f}} & 
\multicolumn{1}{c}{$N_{\rm H}$} & 
\multicolumn{1}{c}{$T_{\rm in}$} & \multicolumn{1}{c}{$R_{\rm in}$\tablenotemark{c}} & 
\multicolumn{1}{c}{$\chi^2_{\nu}$/dof} \\
\multicolumn{1}{c}{~} & \multicolumn{1}{c}{(10$^{22}$\,cm$^{-2}$)} &
\multicolumn{1}{c}{(keV)} & \multicolumn{1}{c}{(km)} & 
\multicolumn{1}{c}{(10$^{22}$\,cm$^{-2}$)} & \multicolumn{1}{c}{(keV)} &
\multicolumn{1}{c}{(km)} & \multicolumn{1}{c}{~} \\
\tableline
GRO\,J1655$-$40 & 2.17\tablenotemark{d} & 0.60$\pm$0.04 & 34$\pm$5 & 79$^{+49}_{-27}$ & 1.1$\pm$0.1 & 15$^{+31}_{-5}$ & 1.4/41 \\
4U\,1630$-$47 & 12.3\tablenotemark{d} & 0.60$\pm$0.03 & 49$\pm$6 & 76$^{+37}_{-30}$ & 1.4$\pm$0.2 & 12$^{+14}_{-4}$ & 1.6/41 \\
\tableline
\multicolumn{1}{l}{~} & 
\multicolumn{7}{c}{partial covering absorption of DBB\tablenotemark{g}} \\
\multicolumn{1}{c}{dip} & 
\multicolumn{1}{c}{$N_{\rm H_{\rm int}}$} & \multicolumn{1}{c}{~} & 
\multicolumn{1}{c}{$F$} & 
\multicolumn{1}{c}{$N_{\rm H}$} & 
\multicolumn{1}{c}{$T_{\rm in}$} & \multicolumn{1}{c}{$R_{\rm in}$} &
\multicolumn{1}{c}{$\chi^2_{\nu}$/dof} \\
\multicolumn{1}{c}{~} & 
\multicolumn{1}{c}{(10$^{22}$\,cm$^{-2}$)} & 
\multicolumn{1}{c}{~} & \multicolumn{1}{c}{~} & 
\multicolumn{1}{c}{(10$^{22}$\,cm$^{-2}$)} &
\multicolumn{1}{c}{(keV)} & \multicolumn{1}{c}{(km)} & 
\multicolumn{1}{c}{~} \\
\tableline
GRO\,J1655$-$40 & 2.17\tablenotemark{d} & & 0.985$\pm$0.007 & 191$\pm$17 & 0.89$\pm$0.03 & 100$\pm$29 & 1.7/42 \\
4U\,1630$-$47 & 5.7$\pm$1.0 & & 0.97$\pm$0.01 & 299$^{+77}_{-40}$ & 1.19$\pm$0.07 & 67$^{+35}_{-15}$ & 1.2/41 \\
\tableline
\multicolumn{8}{l}{\tablenotemark{a}\,\,Model: $e^{-N_{\rm H_{\rm int}}\sigma(E)}$[A$_{\rm pl}E^{-\gamma }+e^{-N_{\rm H}\sigma (E)}$[DBB($T_{\rm in},R_{\rm in}$)]], where $\sigma (E)$ the photo-electric absorption cross section.}\\
\multicolumn{8}{l}{\tablenotemark{b}\,\,GRO\,J1655$-$40 dip: $\overline{I}$=878\,cts\,s$^{-1}$; 4U\,1630$-$47 dip: $\overline{I}$=860\,cts\,s$^{-1}$.} \\
\multicolumn{8}{l}{\tablenotemark{c}\,\,In the case of 4U\,1630$-$47 in units of ${1\over{\sqrt{\cos i}}}\,\cdot\,{\rm d}_{10}$, see also Table~2.} \\
\multicolumn{8}{l}{\tablenotemark{d}\,\,Parameter fixed to the persistent emission value.} \\
\multicolumn{8}{l}{\tablenotemark{e}\,\,Model: $e^{-N_{\rm H_{\rm int}}\sigma(E)}$[BB($T_{\rm bb}$,$R_{\rm bb})+e^{-N_{\rm H}\sigma (E)}$[DBB($T_{\rm in},R_{\rm in}$)]].} \\
\multicolumn{8}{l}{\tablenotemark{f}\,\,Apparant black-body radius at 3.2\,kpc (GRO\,J1655$-$40) or 10\,kpc (4U\,1630$-$47).} \\
\multicolumn{8}{l}{\tablenotemark{g}\,\,Model: $e^{-N_{\rm H_{\rm int}}\sigma(E)}$[$Fe^{-N_{\rm H}\sigma (E)}+(1-F)$][DBB($T_{\rm in},R_{\rm in}$)], where $F$ is the covering fraction.} \\
\end{tabular}
\end{table}
\normalsize

\newpage

\begin{figure}
\caption{}
\psfig{figure=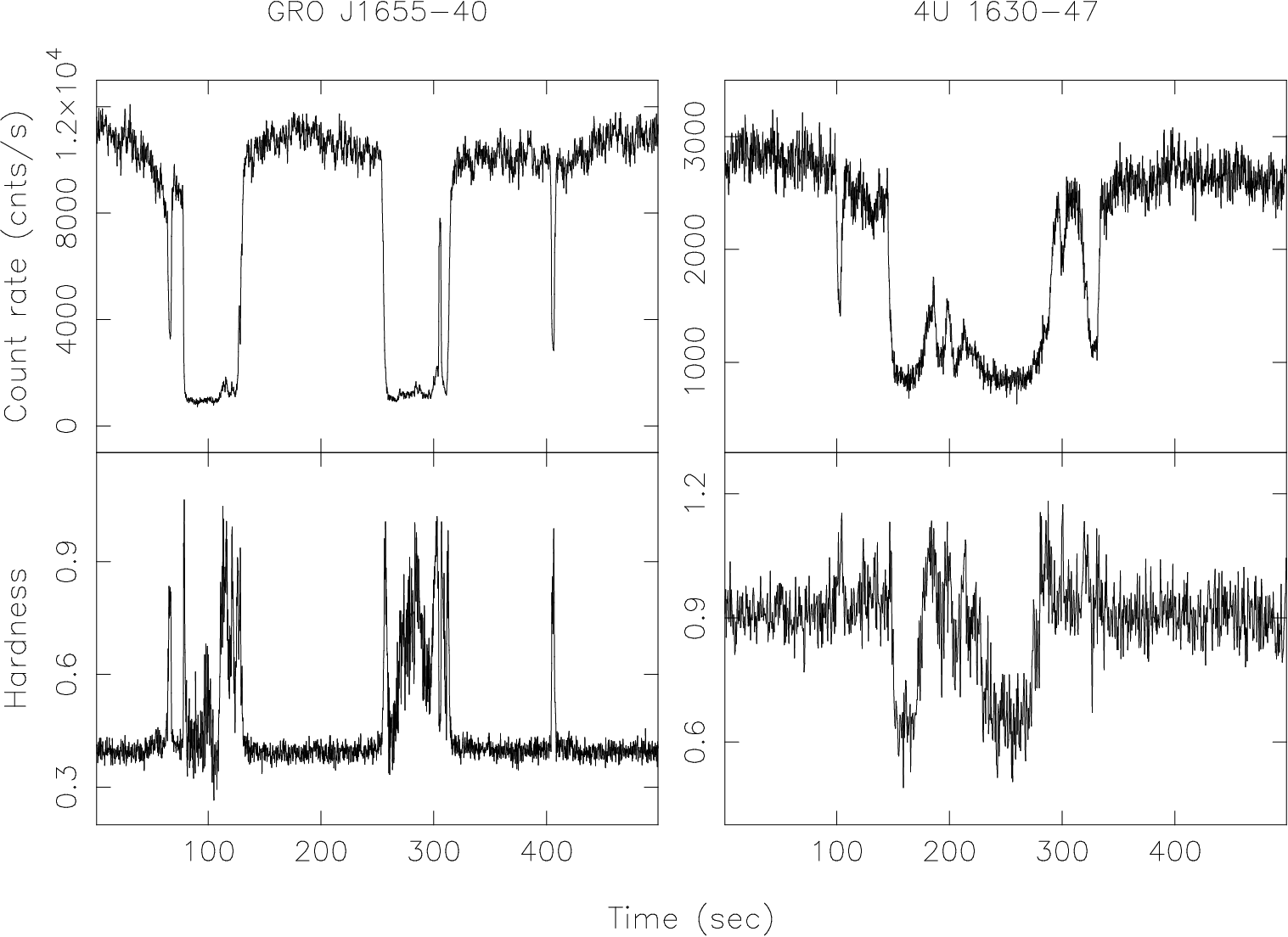,bbllx=33pt,bblly=36pt,bburx=568pt,bbury=769pt,angle=-90,width=16cm}
\end{figure}

\begin{figure}
\caption{}
\psfig{figure=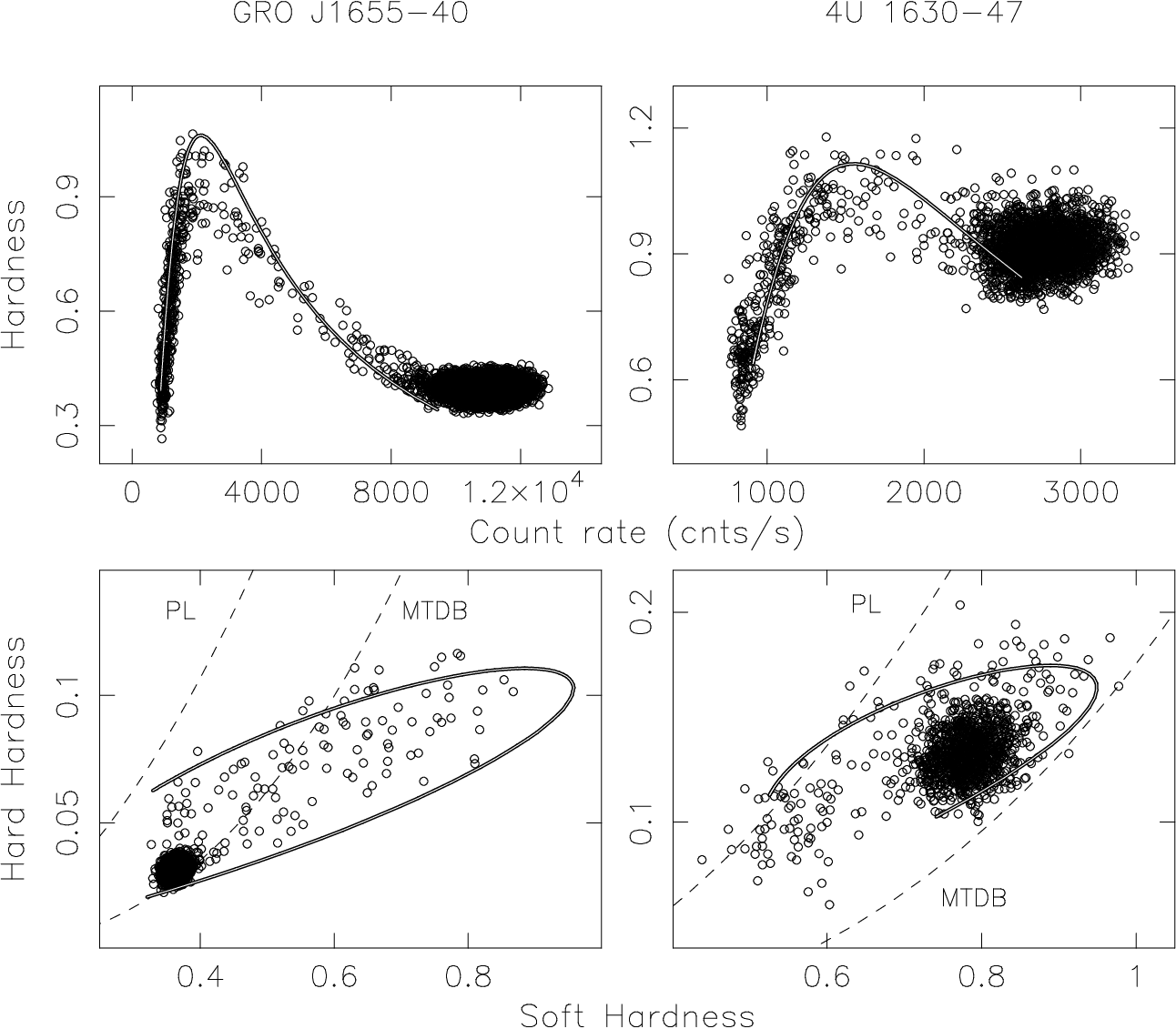,bbllx=28pt,bblly=39pt,bburx=581pt,bbury=664pt,angle=-90,width=14cm}
\end{figure}

\begin{figure}
\caption{}
\psfig{figure=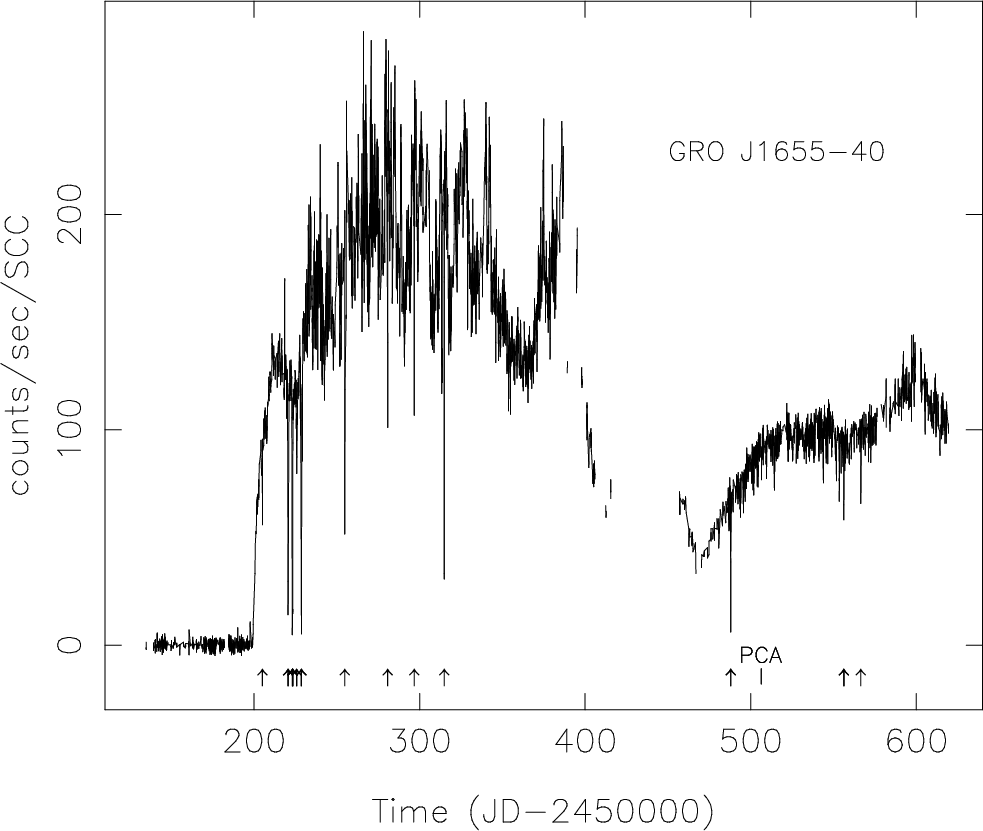,bbllx=56pt,bblly=298pt,bburx=534pt,bbury=702pt,width=12cm}
\end{figure}

\end{document}